\def\eg{$e_g$}

\def\t2g{$t_{2g}$}
\def\lca{La$_{1-x}$Ca$_{x}$MnO$_3$}
\def\lc9{La$_{0.81}$Ca$_{0.19}$MnO$_3$}
\def\lcz{La$_{0.84}$Ca$_{0.16}$MnO$_3$}
\def\lcv{La$_{0.75}$Ca$_{0.25}$MnO$_3$}
\def\2+{$^{2+}$}
\def\3+{$^{3+}$}
\def\4+{$^{4+}$}
\def\ea{\emph{et al.}}
\def\Omn{O$-$Mn$-$O}

\documentclass[twocolumn,superscriptaddress,prb,showpacs]{revtex4}
\usepackage{graphicx}
\begin{document}

\title{Structural response to O*-O' and magnetic transitions in orthorhombic perovskites}

\author{Bas B. \surname{Van Aken}}
\affiliation{Solid State Chemistry Laboratory, Materials Science Centre, University of Groningen, Nijenborgh 4, 9747 AG  Groningen, the Netherlands}
\author{Auke Meetsma}
\affiliation{Solid State Chemistry Laboratory, Materials Science Centre, University of Groningen, Nijenborgh 4, 9747 AG  Groningen, the Netherlands}
\author{Y. Tomioka}
\affiliation{Correlated Electron Research Center (CERC), National Institute of Advanced Industrial Science and Technology (AIST), Tsukuba 305-8562, Japan}
\affiliation{Joint Research Centre for Atom Technology (JRCAT), National Institute of Advanced Industrial Science and Technology (AIST), Tsukuba 305-0046, Japan}
\author{Y. Tokura}
\affiliation{Correlated Electron Research Center (CERC), National Institute of Advanced Industrial Science and Technology (AIST), Tsukuba 305-8562, Japan}
\affiliation{Joint Research Centre for Atom Technology (JRCAT), National Institute of Advanced Industrial Science and Technology (AIST), Tsukuba 305-0046, Japan}
\affiliation{Department of Applied Physics, University of Tokyo,
Bunkyo-ku, Tokyo 113-8656, Japan}
\author{Thomas T. M. Palstra}
\email{palstra@chem.rug.nl}
\affiliation{Solid State Chemistry Laboratory, Materials Science Centre, University of Groningen, Nijenborgh 4, 9747 AG  Groningen, the Netherlands}

\date{\today}

\begin{abstract}
We present a temperature dependent single crystal x-ray diffraction
study of twinned orthorhombic perovskites \lca, for $x=0.16$ and $x=0.25$. These data show the evolution of the crystal structure from the ferromagnetic insulating state to the ferromagnetic metallic state. The data are modelled in space group $Pnma$ with twin relations based on a distribution of the $b$ axis over three perpendicular cubic axes. The twin model allows full structure determination in the presence of up to six twin fractions using the single crystal x-ray diffraction data. 
\end{abstract}


\maketitle

\section{Introduction}
The manganites have generated considerable attention because of the colossal
magnetoresistance effect. The role of magnetic order has been widely discussed
since double exchange, inducing the ferromagnetic order, is required to generate
a metallic ground state. The role of orbital order is much less understood.
While {\it local} Jahn-Teller (JT) distortions are crucial in explaining the
localisation of the charge carriers in the paramagnetic state, many of the structural features of the {\it long range} Jahn-Teller (JT) ordering has not been well studied and warrant closer investigation. Even for undoped systems the JT ordering is not well understood as the electronic degrees of freedom have a strong interaction with strain.\cite{Mat01} The orbital ordering in perovskites with degenerate $e_g$ electrons can be easily measured, whereas for degenerate $t_{2g}$ electrons the Jahn-Teller distortions are much smaller. In this paper we focus on the effects of doping on the JT ordering and the role of the lanthanide site shift. In order to minimise the effect of "extraneous" strain, we have studied high quality single crystals. A method to obtain the crystal structure of twinned orthorhombic samples is presented using single crystal x-ray diffraction (SXD). Furthermore, we will show how the generally observed $Pnma$ symmetry can accommodate a 3D rotation of the octahedra, known as the GdFeO$_3$ rotation, concurrently with a Jahn-Teller ordering.

\subsection{General}
Generally, three basic states for manganites are distinguished, \emph{viz.} the
ferromagnetic metal, the paramagnetic polaronic liquid and the orbital and/or
charge ordered antiferromagnetic insulating state. These states can be
identified by different structural features. Metallicity induces charge
delocalisation and is associated with equal Mn$-$O bond lengths. The charge
and/or orbital ordered phases give rise to, for instance, lattice doubling. The
polaronic liquid is locally characterised by small polarons.

\subsection{diffraction}
Typically crystal structure information is generated by neutron powder diffraction (NPD). NPD gives direct information about the interplanar distances and is very accurate in determining the changes of the lattice parameters with temperature and pressure. Using a full pattern Rietveld Analysis the atomic positions can be determined with an accuracy of $10^{-4}$-$10^{-3}$. NPD is very powerful for oxides because of the large nuclear scattering factor for neutrons of oxygen, compared with the low scattering factor for X-rays. However NPD is not very suitable to determine the symmetry of the crystal structure.

The method that we have used is SXD. SXD is a very powerful experimental application to determine crystal structures of many materials. Whereas powder diffraction projects the reciprocal space to one dimension ($d$~spacing, or diffraction angle $2\theta$), SXD retains the full 3-dimensional information of the reciprocal lattice. The relative co-ordinates of the atoms in the unit cell can be determined from the intensity distribution with an accuracy better than by NPD. 

However, for many perovskites SXD is not used because of the unconventional twinning of the crystals. The structure of orthorhombic perovskites is a deformation of the structure of a simple cubic perovskite. The deviations from cubic symmetry are relatively small. Therefore, domains with specific orientations of the unit cell can grow coherently on each other, a phenomenon known as twinning. Twinning of the perovskites is complex as the distortions from the cubic symmetry allow not only simple twin axes and twin planes, but also 3D twin relations. Then the observed intensities of the diffraction spots in reciprocal space are the weighted sums of the same lattice in different orientations, with a weighting factor proportional to the volume of the corresponding twin fraction. Detwinning is the process of finding the original intensity distribution, allowing a full structure refinement.

A method is presented to analyse the 3D twinned crystals, and to apply
a full structure determination. Besides it is shown that one does not
need a larger data set than for untwinned crystals, which is
unusual. Using SXD we observe a $2a_p\times2a_p\times2a_p$ unit cell,
with $a_p$ the lattice parameter of the simple cubic perovskite. It is
commonly accepted that the unit cell of \lca\ is
$\sqrt{2}a_p\times2a_p\times\sqrt{2}a_p$. The twinning is unique and
it involves a distribution of the $b$~axis over three perpendicular
cubic axes.\cite{Dec00} This leads to a large fraction, about 25\%, of
coinciding reflections and an only twice as large unit cell. The
observed unit cell originates from a three-dimensional type of
twinning that is not restricted to manganites. The model is most
likely of general application for a large variety of perovskite $Pnma$
crystals.

\section{Crystal structure}
Most structure research of perovskites focuses on the Mn$-$O distances and the
Mn$-$O$-$Mn angles, as these parameters determine the competition between the
superexchange and double exchange interactions. The approach here stresses the
importance of a complete structure refinement, including the La-position and the
rotation/distortion of the MnO$_6$ octahedra. The main deformations that
determine the deviation from the cubic structure are the GdFeO$_3$ rotation and
the Jahn-Teller distortion.

The basic building block of the perovskite structure is a $\sim3.9$ {\AA}
cubic unit cell with Mn in the centre and O at the face centres. The oxygen ions
co-ordinate the Mn ions to form MnO$_6$ octahedra. The A~atoms are located at
the corners of the cube. The undistorted 'parent' cubic structure rarely exists,
for instance SrTiO$_3$\cite{von46}, but distorts to orthorhombic or rhombohedral
symmetry.

\subsection{GdFeO$_3$ rotation}
\begin{figure}[htb]
   \centering
   \includegraphics[width=80mm]{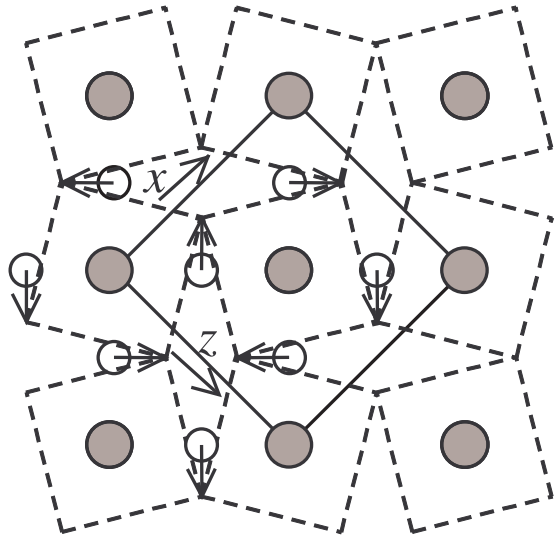}
   \caption{Sketch of the GdFeO$_3$ rotation in the $ac$~plane,
   obeying $Pnma$ symmetry. Mn and O ions are represented by large
   gray and small
   open circles, respectively. Open arrows indicate the movement associated with the
   GdFeO$_3$ rotation. The full line indicates the unit cell.}
   \label{fig:rot}

\end{figure}

Due to the small radius of the A~site ion, with respect to its surrounding cage,
the MnO$_6$ octahedra tilt and buckle to accommodate the lanthanide ion. This is
known as the GdFeO$_3$ rotation\cite{rotation2} and yields the space group
$Pnma$. The cubic state allows one unique oxygen position. Due to the GdFeO$_3$
rotation we need two non-equivalent oxygen positions to describe the structure in $Pnma$ symmetry. O2 is the in-plane oxygen, on a general position, $(x, y, z)$. Two opposite Mn$-$O2 bonds have the same length, but the perpendicular bonds do not need to be equal. O1 is the apical oxygen, located on a fourfold $(x, \frac{1}4, z)$ position on the mirror plane. Mn$-$O1 bonds are always of the same length. Both in the undistorted and the distorted perovskite, the O$-$Mn$-$O bond angles are $180^\circ$ (or near $90^\circ)$, but due to the buckling Mn$-$O$-$Mn bond angles are significantly less than $180^\circ$. A pure GdFeO$_3$ rotation can be obtained within $Pnma$ with equal Mn-O bond lengths, as is seen in Fig.~\ref{fig:rot}.

The GdFeO$_3$ rotation is strongly related to the tolerance factor $t$, which is defined as:

\begin{equation}
  t=\frac{\langle r_{A^{2+/3+}}\rangle+r_{O^{2-}}}{\sqrt2(\langle r_{Mn^{4+/3+}}
  \rangle+r_{O^{2-}})},
  \label{eq:tolerance factor}
\end{equation}

where $r_x$ is the ionic radius of element $x$ and $\langle r\rangle$ denotes
the average radius. For an ideal perovskite structure the ratio between the radii of the A~site ion and the transition metal ion is such that the tolerance factor is equal to one. In the La-Ca system the tolerance factor varies from 0.903 to 0.943 going from LaMnO$_3$ to CaMnO$_3$. For the A~site, a co-ordination number of nine is assumed, values have been taken from Shannon and Prewitt\cite{Shannon}.

The rotation also depends on the temperature. At high temperature all perovskite structures are (near) cubic, due to the thermal motion of the ions. These motions diminish with decreasing temperature, thereby decreasing the 'effective' radius of the ions. As a result, the lanthanides prefer a tighter co-ordination and this is achieved by an increase in the tilting and buckling of the octahedra.

The rotation is also sensitive to a ferromagnetic ordering. A ferromagnetic ordering is usually accompanied by an increase in the molar volume. For the $Pnma$ perovskites, lessening the rotation will yield an increase in molar volume. Therefore, ferromagnetic ordering will be characterised by a step in the rotation parameter, with the smaller rotation in the ferromagnetic phase.

\subsection{Jahn-Teller distortion}
\begin{figure}[htb]
   \centering
   \includegraphics[width=80mm]{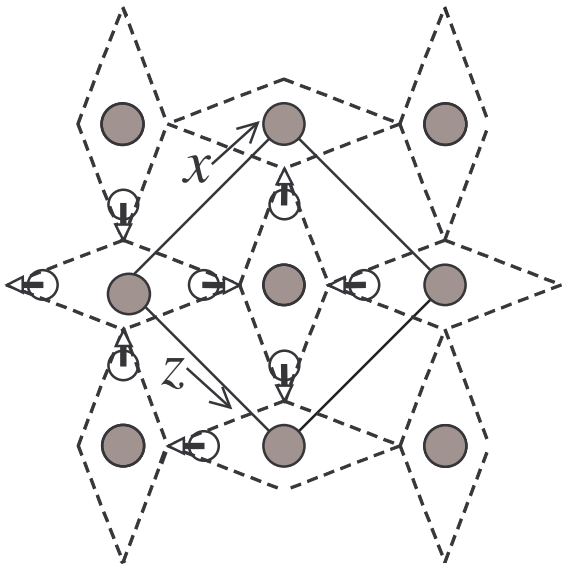}
   \caption{Sketch of the $Q2$ JT distortion in the $ac$~plane, obeying $Pnma$
   symmetry. The arrows indicate the movement associated with the JT distortion. 
   The full line indicates the unit cell.}
   \label{fig:JT}
\end{figure}

The Jahn-Teller effect originates from the degenerate Mn\3+ $d^4$ ion in an octahedral crystal field. Two possible distortions are associated with the Jahn-Teller effect. $Q2$ is an orthorhombic distortion, with the in-plane bonds differentiating in a long and a short one. $Q3$ is the tetragonal distortion with the in-plane bond lengths shortening and the out-of-plane bonds extending, or \emph{vice versa}\cite{Kan60,Yam96}. The main result of the distortions $Q2$ and $Q3$ is that the Mn$-$O distances become different and the degeneracy of the \t2g and \eg\ levels is lifted. In Fig.~\ref{fig:JT}, a $Q2$ distortion is shown.

For LaMnO$_3$, the orbital splitting is such that $z^2$-like orbitals are
occupied, alternately oriented along the $x+z$ and
$x-z$~axis\cite{Goo55,Ele71}, yielding the $Q2$ distortion. The O2 fractional
co-ordinate and this $Q2$ distortion are thus intimately related. For a $Q3$
distortion, only the ratio between the $b$ lattice parameter and the $a$ and $c$
lattice parameters has to change.

\subsection{Glazer's view on the octahedra}
There are three rotational degrees of freedom for a rigid MnO$_6$ octahedron. Glazer
identified all possible sequences of these rotations for the perovskite
system\cite{Gla72}. Two of these sequences create the symmetry elements that
make up the standard perovskites. In Glazer's notation: $a^-b^+a^-$\cite{glazer}
yields orthorhombic $Pnma$ and $a^-a^-a^-$ will result in rhombohedral
$R\overline{3}c$. As Glazer worked in a pseudocubic $2a_p\times2a_p\times2a_p$
system we have to transform the rotation $a^-b^+a^-$ to the orthorhombic unit
cell, which gives $a^-b^+c^0$. This means that the rotation around the $c$~axis,
$\eta_z$, is zero. Any movement of an oxygen
position will also affect the position of the La atom\cite{Miz99}. 

\begin{figure}[htb]
   \centering
   \includegraphics[bb=25 645 170 820]{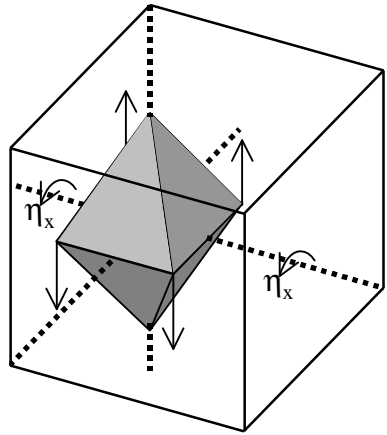}
   \caption{Rotation, $\eta_{x}$, of the MnO$_6$ octahedron around the
   $a$~axis. The arrows indicate the motion of the O2 ions.}
   \label{fig:glazer}
\end{figure}

By rotating the MnO$_6$ octahedra only around the $x$~axis, two oxygen ions are
raised from their cubic position and two are lowered as is sketched in
Fig.~\ref{fig:glazer}. Here we neglect the influence of a rotation $\eta_y$ on
these co-ordinates. This is in agreement with the observation that the four
positions of O2 near the Mn at (0, 0, 0) have the same distance to the $y=0$
plane, albeit two are positive and two are negative. As a result of $\eta_z=0$
the fractional $x$~co-ordinate of O1 is zero. In practice small deviations
are found, indicative of the non-rigid behaviour of the MnO$_6$ octahedra.
Typically in AMnO$_3$ these two rotations and additionally the $Q2$ JT distortion are observed.

Another effect of the GdFeO$_3$ rotation is that the number of formula units per
unit cell is enlarged. The unit cell is doubled in the $b$~direction, with
respect to the original cubic cell. The $ac$~plane is also doubled, resulting in
$b\approx2a_p$, and $a\approx c\approx \sqrt{2}a_p$. Despite the rotations and
distortion, the structure remains in origin cubic. The shifts of the atoms in
the unit cell are small. Furthermore, Mn remains octahedrally surrounded although
the Mn$-$O distances tend to differ and some \Omn\ angles may no longer be
perfectly $90^{\circ}$. The $180^{\circ}$ \Omn\ bond angles are intrinsic to the
inversion centre at the Mn position.

All oxygen positions are related by the symmetry operations of the space group
$Pnma$. That means that the in-plane co-ordinates of one ion ($x_{O2},z_{O2}$ ) determine
all other in-plane positions. However, any linear combination of these two
parameters and its orthogonal complementary combination will be as good.
Normally we prefer $x_{O2}$ and $z_{O2}$, since we refine the atomic positions as $x_{O2}$ and $z_{O2}$.
However, the oxygen shifts (arrows) in Fig.~\ref{fig:rot}, indicating the
rotation, can be expressed as a movement along [101]\cite{position}.
We note that these are perpendicular to the oxygen shifts resulting from JT ordering shown in Fig.~\ref{fig:JT}.

\begin{figure}[htb]
   \centering
   \includegraphics{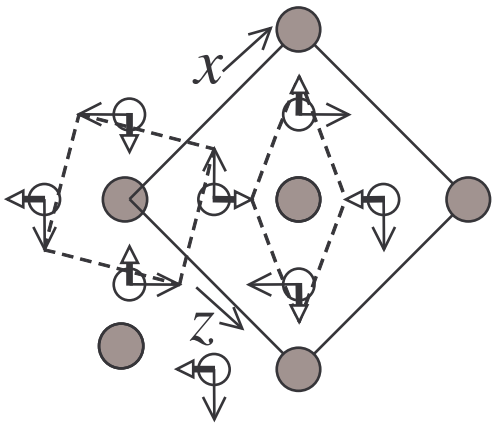}
   \caption{Sketch of the GdFeO$_3$ rotation (open arrow) and the JT
   distortion ($\Delta$-arrow) in the \emph{ac}~plane, obeying \emph{Pnma}
   symmetry. Mn and O are represented by large gray and small open circles,
   respectively.}
   \label{fig:plane}
\end{figure}

Thus,  we can express  the position  of the  oxygen atoms  relative to
their "cubic"  position by the 'shift'  parameters $x_{O2}+z_{O2}$ and
$x_{O2}-z_{O2}$ as sketched  in Fig.~\ref{fig:plane}. Here, a movement
of  the O2 ion  parallel to  [101]\cite{position}, which  implies that
$x_{O2}+z_{O2}=\frac{1}{2}$,  results  in  equal  bond lengths  but  a
Mn$-$O$-$Mn angle smaller than 180{$^\circ$}.  We interpret this
movement, along [101], as the GdFeO$_3$ rotation. 
Similarly, a shift of the O~ion along [10$\overline{1}$], thereby fixing
$x_{O2}-z_{O2}=0$,  results  in different  in-plane  bond lengths  and
therefore  indicates  a Jahn-Teller  distortion.  The $Q2$  distortion
variable is defined as $x_{O2}+z_{O2}-\frac{1}{2}$, thus no distortion
yields  a $Q2$  distortion parameter  equal to  zero.  The undistorted
cubic   structure   obeys   both   $x_{O2}+z_{O2}-\frac{1}{2}=0$   and
$x_{O2}-z_{O2}=0$. It is  clear that any O2 position  can be expressed
uniquely  as  a  superposition  of  a  GdFeO$_3$  rotation  parameter,
\emph{viz.}   $x_{O2}-z_{O2}$,   and   a  JT   distortion   parameter,
\emph{viz.}  $x_{O2}+z_{O2}-\frac{1}{2}$.

\section{structure ABO$_3$ vs ionic radius}
\begin{figure}[htb]
   \includegraphics[width=85mm]{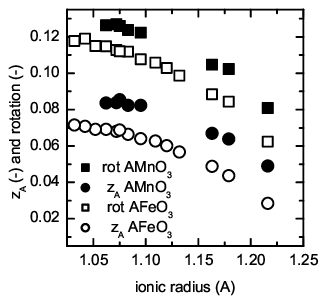}
   \includegraphics[width=85mm]{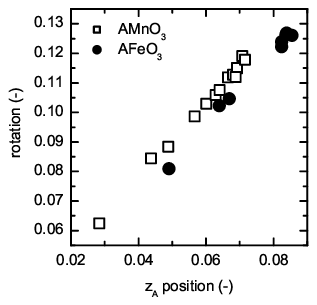}
   \caption{Top panel: The rotation and $z_{\textrm{A}}$ versus the ionic radius
   of AFeO$_3$ and AMnO$_3$. Both parameters indicate a deviation from the cubic
   perovskite, and both increase with decreasing $r_{\textrm{A}}$. Bottom panel:
   The rotation versus $z_{\textrm{A}}$. The parameters show almost perfect
   correlation indicating their intimate relation. Data have been taken from
   ref. [\onlinecite{Mar70}] (AFeO$_3$) and ref. [\onlinecite{Alo00}] (AMnO$_3$). Open and
   closed symbols represent AMnO$_3$ and AFeO$_3$, respectively. Squares and
   circles represent the rotation and site shift parameter, respectively.}
   \label{fig:rot_zet_AFeO3}
\end{figure}

The method of rewriting the oxygen position to two variables, which indicate the
magnitude of the JT ordering and the magnitude of the GdFeO$_3$ rotation, is
applied to two well-studied series of compounds. Marezio \ea\ have studied the
structure of the AFeO$_3$ single crystals with SXD in great detail.\cite{Mar70} Their data
allow us to focus on the correlation between the ionic radius, $r_A$ and
the magnitude of the rotation. In the top panel of Fig.~\ref{fig:rot_zet_AFeO3},
we see that with decreasing $r_A$ the rotation (open squares) increases
monotonously. We also observe an increasing shift from the cubic position for
the A site atom with decreasing $r_A$.

The basic argument favouring the A~site shift is that the band structure energy
can be optimised by decreasing the A$-$O distances for the three shortest bonds
and increasing the A$-$O bond lengths for the three next-shortest bonds. This
effect is demonstrated by the cubic-tetragonal transition in WO$_3$\cite{DeW99}.
The tungsten ion is octahedrally surrounded by oxygen ions. Band structure
calculations show that the gap is lowered by a ferroelectric displacement of the
W ion off-centre. The energy is lowered by the overlap between the empty W $5d_{xz}$ and $5d_{yz}$ orbitals with the occupied O $2p_x$ and $2p_y$,
respectively\cite{DeW99}.

Goodenough showed the importance of interactions between the A~ion, $d^0$, and
the filled oxygen $2p$ orbitals in the case of a GdFeO$_3$
rotation\cite{Goo63,Goo71}. Recently, Mizokawa \ea\ reported on the interplay
between the GdFeO$_3$ rotation, the orbital ordering and the A~site shift in
ABO$_3$, with B=Mn\3+ ($3d^4$) or V\3+ ($3d^2$). Their theoretical calculations
suggest that the observed orbital ordering in LaMnO$_3$ is stabilised by both a
large GdFeO$_3$ rotation and a shift of the A~site ion \cite{Miz99}. Conversely,
if a Jahn-Teller distortion is present, then the energy will be lowered if it is
accompanied by a shift of the A~site.

In cubic perovskites, the A~ion is in the centre of the AO$_{12}$ polyhedra. The
GdFeO$_3$ rotation distorts the AO$_{12}$ polyhedron. Additionally, 
minimising the bonding-antibonding energy will decrease the shortest
A$-$O bonds even more by a shift of the A~ion. Marezio \ea 's data show the
correlation between the rotation and the La~site shift \cite{Mar70}. The
relevant atomic parameters, $x$ and $z$ (open circles) of the A~site, and the
rotation parameter (open squares) are fully correlated. The bottom panel shows a perfect linear
relation between the A~site shift and the rotation parameter of AFeO$_3$ (open
symbols). Thus the shift of the A~site atom is, in this system, fully determined
by the rotation of the FeO$_6$ octahedron. This conclusion is supported by
inspection of the neutron powder diffraction data on AMnO$_3$ (closed symbols)
by Alonso \ea.\cite{Alo00} Their data show that the JT distortion parameter is
$\sim0.0301(8)$ at $T=295$~K, independent of the A ion radius\cite{Alo00}. The
observed slope of rotation \emph{vs.} site shift is identical for AMnO$_3$ and
AFeO$_3$ within the accuracy of the measurement.

The A~site shift is the result of the covalent interaction between the A\3+
ion and the surrounding oxygen ions. The data for AFeO$_3$ shows that the ratio
${x_A}/{z_A}=5$. Thus the A ions will decrease their distance to the
nearest three oxygen ions by moving along a vector ~[1 0 5]. As the A
co-ordination is directly determined by the changes in rotation and distortion,
the A site shift strongly depends on both the GdFeO$_3$ rotation and the JT
distortion.

Although thorough structural studies on the manganites are published in the literature, see \emph{e.g.} ref.~\onlinecite{Dab99}, they typically are based on NPD data of La$_{1-x}$Sr$_{x}$MnO$_3$. In 1971, Elemans \ea\ studied the structure of LaMnO$_3$ and La$_{0.95}$Ca$_{0.05}$MnO$_3$, using NPD. They found that the $Q2$ distortion is similar for $T=4.2$ K and $T=298$ K.\cite{Ele71} The Mn-O bond lengths for \lcv\ are reported by Radaelli \ea. They show a reduction in the distortion or bond length disproportion at $T_c$.\cite{Rad96} Recently, Huang \ea\ report the Mn-O bond lengths for La$_{0.85}$Ca$_{0.15}$MnO$_3$, La$_{0.75}$Ca$_{0.25}$MnO$_3$ and La$_{0.67}$Ca$_{0.33}$MnO$_3$ using NPD. The scatter and error bars make any distinct observation hard. A slight decrease in variance of the bond lengths is observed near $T_c$.\cite{Hua98} In a Raman study, the MnO$_6$ distortion was reported to decrease linearly from 42\% at $x=0$ to 8\% at $x=0.4$, using x-ray powder diffraction.\cite{Lia99} Dabrowski \ea\ studied the relation between oxygen content, doping concentration and crystal structure for \lca. They report a transition at room temperature from the O' state, coherent JT distortions, to the O* state at $x=0.14$.\cite{Dab99b} No sign of coherent JT distortions was found for La$_{0.8}$Ca$_{0.2}$MnO$_3$ and La$_{0.7}$Ca$_{0.3}$MnO$_3$ down to $T=15$ K. Increasing the temperature towards $T_c$, increases the variance in bond lengths, as was determined from total correlation function experiments. This however measures the \emph{local} JT distortions.\cite{Hib99}

\section{experimental procedures}
The experiments were carried out on single crystals of La$_{1-x}$Ca$_x$MnO$_3$, $x=0.16$ and $x=0.25$, obtained by the floating zone method at the JRCAT institute, Tsukuba, Japan. Although all crystals were twinned, small mosaicity and sharp diffraction spots were observed. Furthermore, sharp magnetic and electronic transitions indicate the good quality of the crystals. A thin piece was cut from the crystal to be used for single crystal diffractometry. The single crystal was mounted on an ENRAF-NONIUS CAD4 single crystal diffractometer. The temperature of the crystal was controlled by heating a constant nitrogen flow. Initial measurements were done at 180~K. Temperature dependent measurements between 90~K and 300~K were performed on a Bruker APEX diffractometer with an adjustable temperature set-up. Lattice parameters based on observations by the CCD camera are refined using all observed reflections. Details of the SXD experiments and refinements are published elsewhere.\cite{VanAken} The single crystals exhibit sharp magnetic transitions with $T_c\sim160$ K and 200 K, for \lcz\ and \lcv respectively.

\section{twinning}
\subsection{The twin structure for manganites}
We will focus in this section on the main reason why SXD has not been widely
used for these perovskites, namely twinning. Twinning in doped LaMnO$_3$
originates from the transition of the highly symmetric cubic parent structure to
the orthorhombic symmetry, that accommodates both the GdFeO$_3$ rotation and the
Jahn-Teller distortion. The solution of the twin relations in the crystals
allows us to study in detail the ordering of these compounds as influenced by
temperature, magnetic state and doping concentration. But for now we will focus
on the twin relations.

Twinning can complicate the structure determination and is often
observed in crystals with a reduced symmetry. The orthorhombic
perovskite LaMnO$_3$ has lower symmetry than cubic
SrTiO$_3$. Conventional twin models keep one characteristic axis
unchanged and form the domains by rotation around that axis.  This is
commonly observed in constrained epitaxial thin films and
pseudo-two-dimensional crystals like
YBa$_2$Cu$_3$O$_{7{-}\delta}$. Another standard twin is inversion
twinning, found in non-centrosymmetric systems. Due to the inversion
twin, pseudo-centrosymmetry is obtained and the net polarisation in
ferroelectric compounds is reduced to zero\cite{Rao97a,Van01a}.
Usually, both types of twinning have the property that either all twin
domains have reflections that lie on top of each other, merohedral
twins, or there is a non-commensurate set of reflections, {\it e.g.}
non-merohedral twins. Here a more extensive form of twinning is
proposed, where at a quarter of the observed reflections the
reflections of different domains coincide and the other part of the
observed reflections originates from a single domain only. Together
they give rise to the observed metrically cubic lattice.

Rodr\'{i}gruez-Carvajal \ea\ noted that their neutron powder spectra of pure
LaMnO$_3$ could be indexed in a cubic $2a_p\times2a_p\times2a_p$ unit cell,
above the Jahn-Teller transition temperature\cite{Rod98}. They did not observe
an orthorhombic splitting of the peaks. Nevertheless, they could not refine the
structure in a cubic unit cell. But full pattern refinement was only possible in
the orthorhombic space group $Pnma$\cite{Rod98}. Note that in powder diffraction
data there is no direct method to observe the threefold symmetry along the body
diagonal. The refinement has been done in the conventional $Pnma$ setting. They
suggested that the powder is twinned but could not give evidence, as they only
studied powders. Combining our observation, including the double cubic lattice,
the systematic extinctions and the non-cubic intensity distribution, with the
suggestion given by Rodr$\mathrm{\acute{i}}$gruez-Carvajal \ea\ leads to the
conclusion that our single crystals are twinned.

Electron diffraction and microscopy are instrumental to characterise this twinning because of its high spatial resolution. Recently D\'{e}champs \emph{et al.}\cite{Dec00} identified the twin relations of orthorhombic and rhombohedral AMnO$_3$. Their data are interpreted for $Pnma$ symmetry with coherent domains, with the doubled $b$~axis in three orthogonal orientations.

\begin{figure}[htb]
   \centering
   \includegraphics[bb=60 680 300 820, width=65mm]{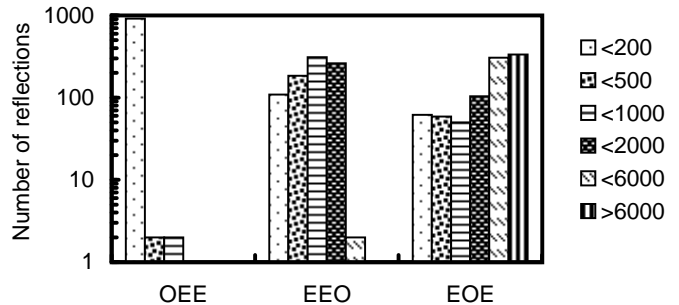}
   \caption{The number of reflections versus the orientations \textsc{oee},
   \textsc{eeo} and \textsc{eoe} for different intensity ranges. \textsc{oee},
   $b'$ parallel to $a''$, has mostly very low intensity. Reflections
   \textsc{eoe}, $b'$ parallel to $b''$, have a broad range of intensities. In
   cubic symmetry, these distributions should be identical for absorption
   corrected data, which we considered.}
   \label{fig:eeo}
\end{figure}

Reflections in reciprocal space are experimentally observed at a regular
distance in three orthogonal directions, corresponding to a cubic lattice
spacing of $7.8$ \AA\ in real space. Although the three axes had equal lengths,
we could not observe the required threefold symmetry axis along
$\langle111\rangle$ for cubic symmetry. This is shown for \textsc{eeo}
reflections in Fig.~\ref{fig:eeo}, \textsc{e}~denotes an even value for the
Miller indices, \textsc{o}~stands for odd. Furthermore, studying the intensity
distribution of $hkl$ planes with constant $h$, $k$ or $l$ showed much
regularity, and some anomalous extinction conditions are observed. An overview
of these features is given in appendix \ref{app:extinctions}. A twinned
structure consisting of coherently oriented $Pnma$ domains is proposed, analogous with the results of D\'{e}champs \emph{et al.}\cite{Dec00}. Due to
partial overlap this results in a metric cubic system with
$a\approx2a_p\approx7.8$~\AA.

\subsection{Twin model}
\begin{figure}[htb]
   \includegraphics[bb=30 620 190 805]{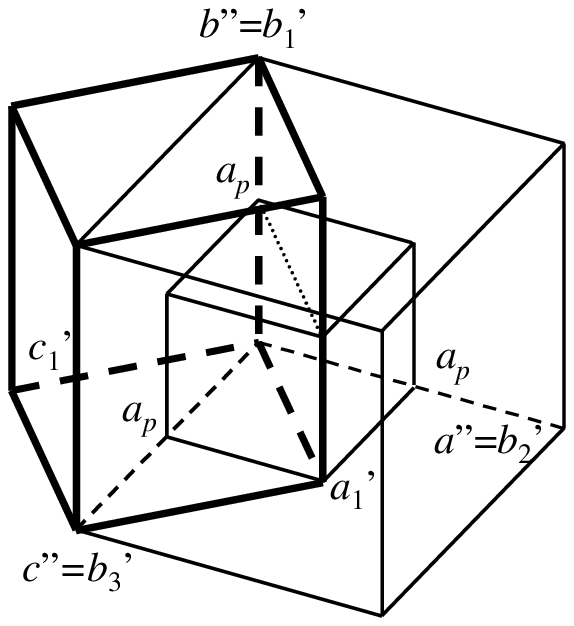}
   \includegraphics[bb=25 645 200 820]{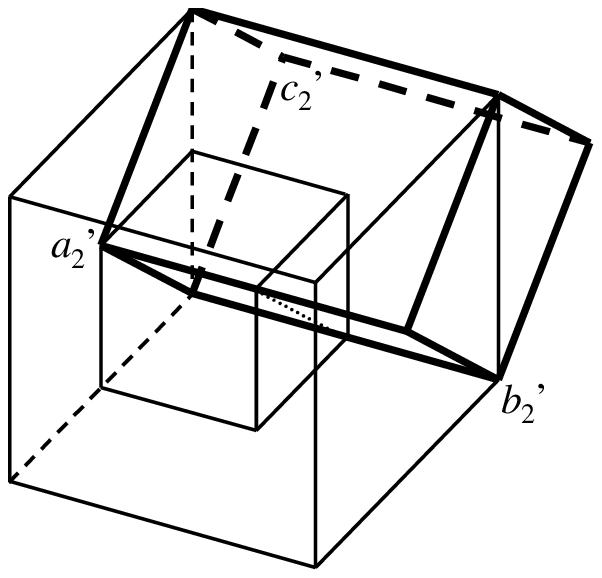}
   \caption{Two of three possible orientations of the $Pnma$ unit cell in the
   observed $2a_{\textrm{p}}\times2a_{\textrm{p}}\times2a_{\textrm{p}}$
   $(a''\times b'' \times c'')$ unit cell. $b'_1$, $b'_2$ and $b'_3$ denote
   the orientations of the doubled axes along the original cubic axes,
   $a_{\textrm{p}}$. Note that the choice of $a'$ and $c'$ is still open, after
   choosing an orientation for the $b'$ axis.}
   \label{fig:2of3}
\end{figure}

The transformation of cubic to orthorhombic
symmetry requires a designation of $a$, $b$ and $c$ with respect to the
degenerate cubic axes. There are three possibilities to position the doubled
$b$~axis along the three original cubic axes. Thus we propose that the three
fractions' $b$~axes are oriented along the three original axes of the cubic unit
cell, as sketched in Fig.~\ref{fig:2of3}. This twin model consists of a $Pnma$
unit cell, transformed by rotation about the 'cubic' [111] axis. We still have
the freedom to choose the $a$ and $c$~axes, perpendicular to the $b$~axis, and
rotated 45$^\circ$ with respect to the cubic axes. Therefore, this model yields
six different orientations of the orthorhombic unit cell. As the differences
between $a/\sqrt{2}$, $b/2$ and $c/\sqrt{2}$ are small, we observe the
reciprocal superposition of the six orientations as metrically cubic.

Using the proposed model, the different contributions of the twin fractions to
the total integrated peak intensity are taken into account, according to the
following procedure. We considered for each observed reflection the corresponding
orthorhombic reflections for the twin fractions (if applicable) and used an
identical unit cell for each twin. The refined model consisted of the regular
parameters in single crystal diffractometer. This way, both the atomic positions
of the asymmetric unit and the volume ratio of the twin domains are refined
simultaneously.

Several crystals were measured and refined with this detwinning model. The
refinements showed that for every crystal the distribution of the volume over
the twin fractions is different. This is another indication that our crystals
are twinned. Other structural deformations than twinning would lead to a
constant, not sample dependent, effect on the structure factors. Attempts to
refine will then result in the wrong model, giving the same value for the volume
fractions, independent of sample. We can also conclude that the size of the twin
domains must be smaller than the magnitude of the measured crystals, \emph{i.e.}
0.1-0.2 mm. Larger twin domains would give rise to crystals of one single domain
and even smaller domains are more likely to produce a constant spreading of the
domains. Non-regular, though constant, volume fraction of the twin domains could
also signal a preferential growth direction of the crystal.

\section{results}
\subsection{Lattice parameters}
\begin{figure}[htb]
   \centering
   \includegraphics[width=80mm]{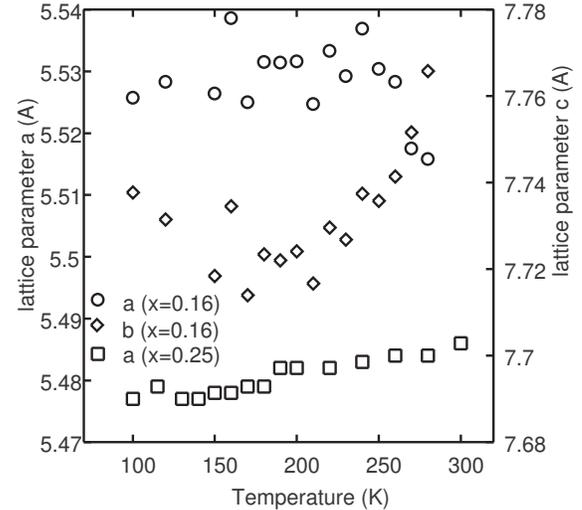}
   \caption{The temperature dependence of the observed lattice parameters $a$. For \lcz\ lattice parameter $b$ is plotted on the right axis. The error bars are of the same size as the symbols. Note that the error bars indicate an under limit to the uncertainty.}
   \label{fig:latpar}
\end{figure}

Reflections in reciprocal space are observed on a double cubic lattice. From these lattice spacings, only the 'average' lattice parameters can be calculated. In Fig.~\ref{fig:latpar} the lattice parameter $a$ is plotted against temperature. For \lcz\ the volume fractions of two of the three $b$ axis orientations were very small, 
therefore, effectively a normal pattern is observed, within only the $a$-$c/c$-$a$ twin. For this sample both lattice parameter $a$ and $b$ are plotted. Lattice parameter $c$ behaved identical to lattice parameter $a$.

The lattice parameter for \lcv\ shows a continuous increase with temperature. No anomaly near $T_c$ can be observed. Lattice parameters $a$ and $b$ for \lcv\ respectively show a sharp down turn and up turn above $T\sim250$~K.

\subsection{Rotation}
\begin{figure}[htb]
   \centering
   \includegraphics[width=80mm]{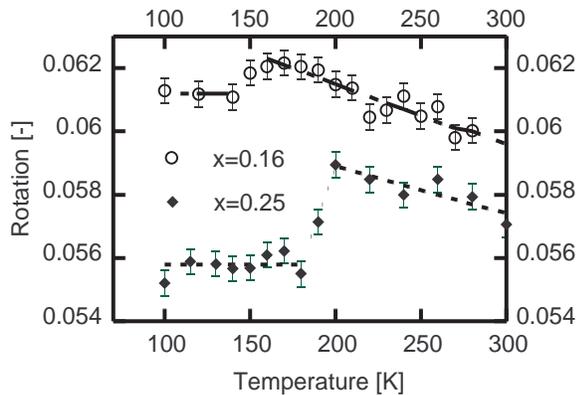}
   \caption{The temperature dependence of the rotation parameter. The broken lines are guides to the eye.}
   \label{fig:rotation 3x vs t}
\end{figure}

In Fig.~\ref{fig:rotation 3x vs t} the temperature dependence of the rotation is
plotted versus the temperature. The rotation-temperature curves depend strongly
on the Ca concentration. The curve of the $x=0.25$ sample shows a sharp steplike increase in the rotation at $T_c$. At $x=0.16$, the scatter in the data is larger, though a smaller step near $T_c$ can still be observed. Above $T_c$, the rotation decreases with a similar slope as the curve for $x=0.25$. The absolute value of the rotation is about 10\% larger for the lower doped sample.

\subsection{JT distortion parameter}
\begin{figure}[htb]
   \centering
   \includegraphics[width=80mm]{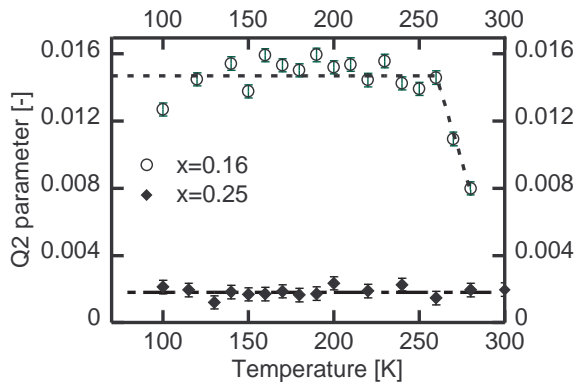}
   \caption{Jahn-Teller parameter against temperature. The broken lines are
   guides to the eye.}
   \label{fig:JT vs t}
\end{figure}

Fig.~\ref{fig:JT vs t} shows that the sample with $x=0.16$ has a constant value
for the JT distortion parameter at low temperature. Above $T=260$~K the JT
distortion parameter decreases rapidly with temperature, which signals the JT orbital ordering temperature.

The JT distortion parameter for the $x=0.25$ sample is independent on
temperature within the uncertainty of the experiment. The value of 0.002 results from the fact that the O2 position is not constrained by symmetry.

\subsection{La~site shift}
\begin{figure}[htb]
   \centering
   \includegraphics[width=80mm]{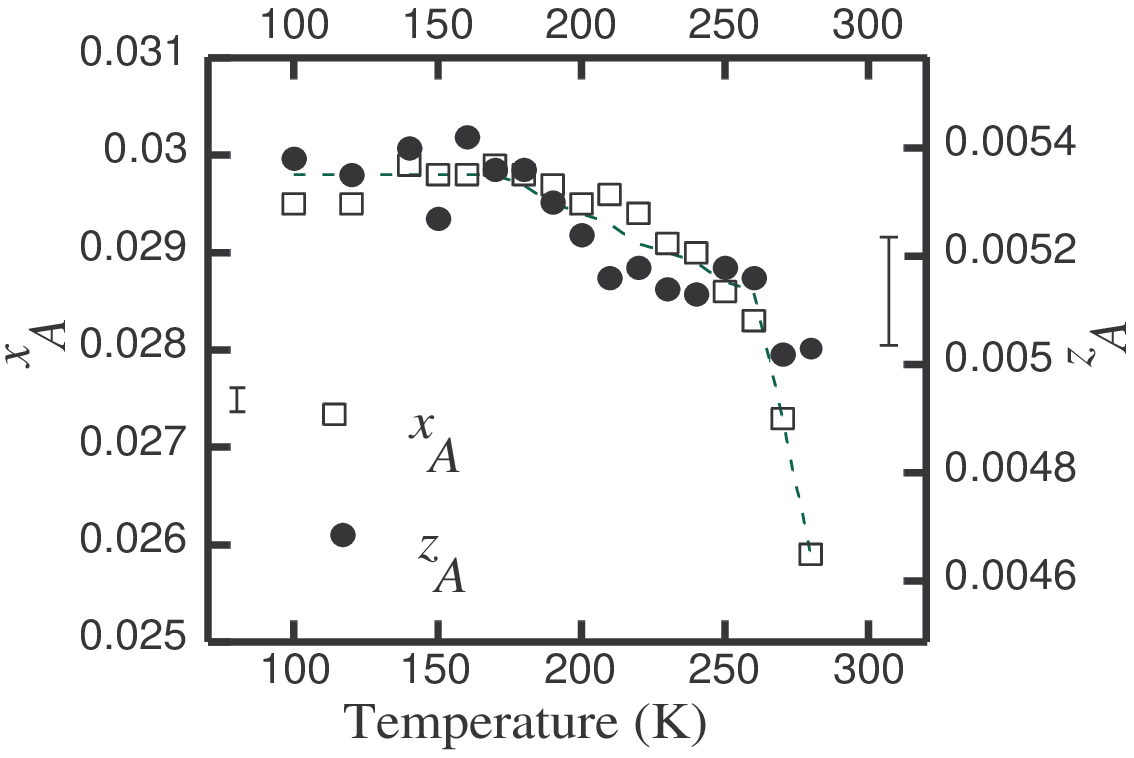}
   \includegraphics[width=80mm]{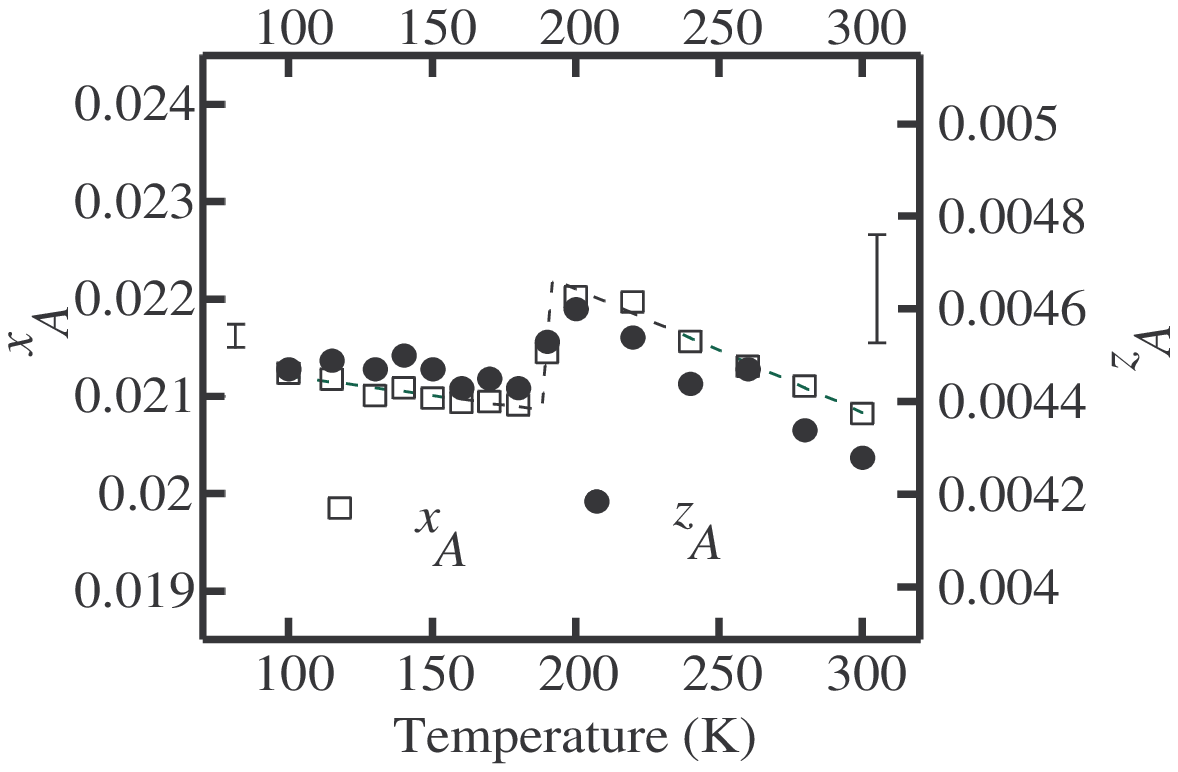}
   \caption{The temperature dependence of the A site position of \lcz\ (top) and
   \lcv\ (bottom). The ratio between $x_A$ and $z_A$ is 5.6 and 4.8. The $x_A$ axes have the same scale to allow comparison. The corresponding $z_A$ axes have been scaled with the average ratios 5.6 and 4.8.}
   \label{fig:shift vs t}
\end{figure}

In Fig.~\ref{fig:shift vs t}, we show the temperature dependence of the A~position, ($x_{A}, \frac{1}{4}, \frac{1}{2}-z_{A}$), with respect to the ideal position ($0, \frac{1}{4}, \frac{1}{2}$). We note that the error bars of the A~position, 0.0001, and scatter are much smaller than the error bars on the JT distortion parameter and rotation parameter, 0.0005. These latter are derived from the oxygen positions, which have a larger uncertainty due to their smaller electron density.

The ratio of $x_A$ and $z_A$ is constant, with $x_A$ roughly 5 times larger than
$z_A$ for all temperatures. 

\section{discussion}
\subsection{Lattice Parameters}
The occurrence of twins in our single crystals effectively averages the lattice parameters $a$, $b$ and $c$, since reflections with different $hkl$, originating from different twin fractions, coincide in one reflection spot. However, for the $x=0.16$ sample, only twin fractions with parallel $b$ axes are observed, which allows a separate determination of $a$/$c$ and $b$. The lattice parameter $c$ is still "averaged" with $a$. 

The lattice parameters, or volume of the unit cell, are influenced by several effects. The most obvious in a temperature dependent study is the thermal expansion. We clearly observe the continuous increase in lattice parameter $a$ with temperature for $x=0.25$. The GdFeO$_3$ rotation is another non-volume conserving distortion of the lattice. The main reason the MnO$_6$ octahedra are rotated is the ionic radius of the A site atom. We expect a change of volume at $T_c$, but we could not observe an anomaly in the lattice parameters at $T=200$~K ($x=0.25$) or at $T=160$~K ($x=0.16$). 

An orbital ordering can have an effect on the lattice parameters, even though it is volume conserving. The effect is different for the $Q2$ and the $Q3$ type distortion. The coherent $Q2$ distortion does not change the lattice parameters in first approximation as the long and short Mn-O bonds are oriented alternately, both in the [101] and the [10$\overline{1}$] direction. However, the $Q3$ distortion has a significant effect on the lattice parameters. Because it has an opposite effect on the Mn-O1 bonds, parallel to $b$, and the Mn-O2 bonds, in the $ac$ plane.

Note that there is a large scatter in Fig.~\ref{fig:latpar}. Combined with the averaging effect of the twin domains, the responses of the crystal structure to the orbital ordering or the ferromagnetic transitions are not very clear. This stresses the importance of an alternative way to study the relation between the structure and the electronic properties.

\subsection{Rotation}
The rotation parameter is clearly sensitive to the ferromagnetic ordering in the
$x=0.25$ samples at $T_c=200$~K. Likewise a smaller transition is shown in the $x=0.16$ curve in Fig.~\ref{fig:rotation 3x vs t}. The decrease in rotation parameter indicates a straightening of the Mn$-$O$-$Mn bond angles as shown in Fig.~\ref{fig:rot} and the text. If a constant volume of the MnO$_6$ octahedra is assumed, the decrease in rotation, associated with the paramagnetic to ferromagnetic transition, yields an increase in total volume. An increase in molar volume, or an increase in the lattice parameters is a common feature of ferromagnetic ordering, as observed for instance in pure nickel.\cite{Wil34} This agrees very well with our observation of the decreasing rotation parameter and consequently increasing volume of the unit cell.

The rotation for LaMnO$_3$ is $0.0810(2)$ at room temperature and decreases with temperature.\cite{Rod98} Elemans \ea\ reported no significant change in rotation between room temperature and 4.2 K, but their resolution is smaller. However, by doping with 5\% Ca a decrease in the rotation parameter to 0.077(1) is observed.\cite{Ele71} CaMnO$_3$ is (almost) cubic and consequently has no rotation. Therefore we expect a continuous decrease in the rotation with increasing doping. This agrees well with our data, which show a decrease of the absolute value of the rotation with 5-10\% between $x=0.16$ and $x=0.25$.

\subsection{JT distortion parameter}
In Fig.~\ref{fig:JT vs t}, we observe that the $Q2$ distortion for the $x=0.25$ sample is constant with temperature, but not equal to zero as expected for the ferromagnetic metallic phase. The value, 0.002(1), is in good agreement with values calculated from structure determinations of AFeO$_3$, \emph{e.g.} 0.0036 for LuFeO$_3$.\cite{Mar70} Fe $3d^5$ in the high spin state, has a completely filled subband, and is therefore not sensitive to the JT effect. However, other effects, for example related to the hybridisation associated with the A site shift, and the degrees of freedom of the O2 position in $Pnma$ symmetry allow the $Q2$ distortion to be non-zero.

The reduction in $Q2$ parameter, as shown by the curve for $x=0.16$, clearly coincides with the JT ordering temperature. As single crystal diffraction is sensitive to long distance correlations, this either means that the orbitals becomes disordered or that there is an overall reduction in the magnitude of the local $Q2$ distortion. Local probes, like pair distribution functions, show that even in the paramagnetic phase the local distortions do not disappear.\cite{Bil96} Our observation of a decrease in $Q2$ distortion is therefore caused by a decrease in the coherence of the orbital order. 

The data for LaMnO$_3$ show that the JT distortion parameter is about 0.0320(10) at room temperature and that the JT ordering reduces to 0.0060 at $T=798$~K, where the orbital order is known to disappear.\cite{Rod98,Mur98b} The low temperature JT distortion parameter and the JT transition temperature strongly depend on the doping level, $x$. With increasing doping both variables decrease.

\subsection{La~site shift}
The ratio of $x_A$ and $z_A$ corresponds to a shift of the A atom along [1 0 5]. Mizokawa \ea assumed the A~site shift along [1 0 7].\cite{Miz99} Their assumed [1 0 7] is in good agreement with our experimental observation of [1 0 5]. 

In fig.~\ref{fig:shift vs t}, changes in the gradient are observed near $T_c$ and the orbital ordering temperature, corresponding with the step in the resistance curve. This supports the assumption that the A~site shift is sensible to changes in the oxygen environment, by changes in the GdFeO$_3$ rotation and the JT distortion.

The A site shift, expressed by $x_A$, in LaMnO$_3$ is 0.0490(2) at room temperature and 0.0217(3) at 798 K\cite{Rod98}. The latter value is comparable with the observed values near 300~K for $x=0.25$.

If we compare these curves with the curves in Fig.~\ref{fig:rotation 3x vs t}
and Fig.~\ref{fig:JT vs t} it is obvious that the curve of rotation \emph{vs.}
temperature of the $x=0.25$ sample is mimicked by the A~site shift \emph{vs.}
temperature curve. Similarly, the A~site shift \emph{vs.} temperature curve of
the $x=0.16$ sample shows the same rapid fall above $T=260$ K as the curve of
the JT distortion parameter \emph{vs.} temperature. But a small change in
slope is observed at $T_c$. Evidently the A site shift is a good probe to pinpoint transitions which involve changes in the oxygen co-ordination of the Mn ions.

\section{conclusion}
We conclude that full structure determination using single crystal x-ray diffraction provides accurate information regarding the magnetic and orbital ordering in manganite perovskites. We showed that the twinning of these single crystals is based on a distribution of the doubled $b$ axis over three perpendicular cubic perovskite axes. The shift of the A site atoms is very sensitive to changes in the oxygen environment of the A site. Therefore, the A site shift is an accurate probe for magnetic transitions and orbital ordering, since these involve rotations and deformations of the MnO$_6$ octahedra network. This warrants structure investigations close to the ferromagnetic insulator to ferromagnetic metal transition near $x=0.2$.

\section{acknowledgements}
This work is supported by the Netherlands Foundation for the Fundamental
Research on Matter (FOM). Stimulating discussions with Gerrit Wiegers, Juan
Rodr\'{i}quez-Carvajal, Lou-F\'{e} Feiner, Graeme Blake and Jan de Boer are
gratefully acknowledged.

\appendix
\section{Data analysis}\label{app:Data_analysis}\label{app:extinctions}
\begin{figure}[htb]
   \centering
   \includegraphics[bb=28 620 300 820, width=80mm]{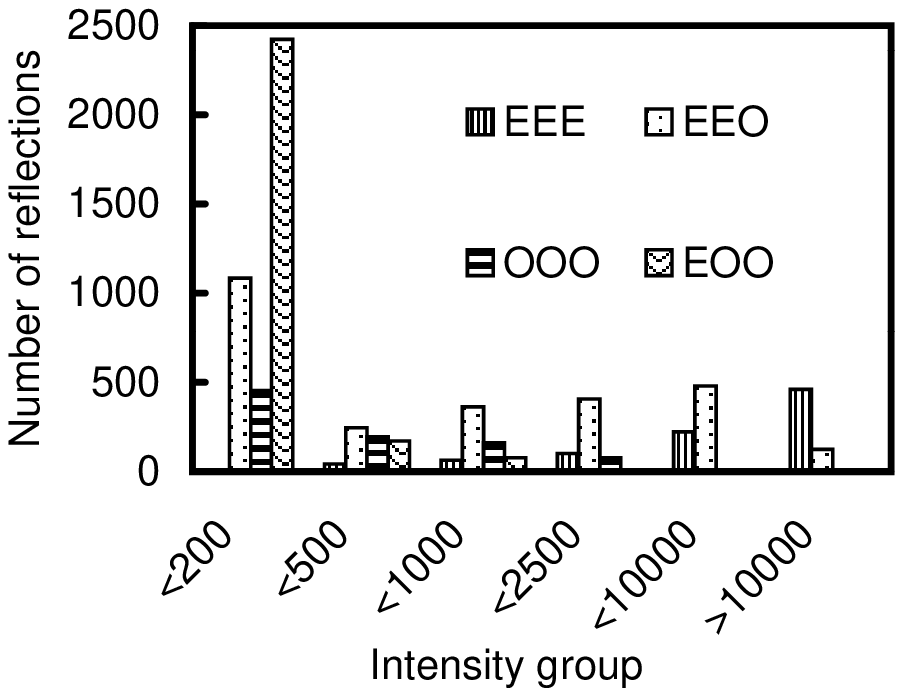}
   \caption{The intensity distribution of reflections. Low intensities,$I<200$ are mostly \textsc{eoo}, while high intensities, $I>2500$, are of the types \textsc{eee} and \textsc{eeo}, \textsc{e} denotes an even value for the Miller indices, \textsc{o} stands for odd. }
   \label{fig:all}
\end{figure}

Although cubic symmetry could not be observed, some regularities and anomalous extinction conditions are present in the intensity distribution over $hkl$ space.  In Fig.~\ref{fig:all} the
observed intensity distribution of the reflections is plotted. The reflections
are subdivided in four groups, on the basis of the number of odd indices in the
double cubic setting. Here the effect of the twin model on the visibility of the
standard $Pnma$ reflection conditions is described and a detailed analysis of
the intensity distribution of a model crystal with $Pnma$ symmetry and of the
measured reflections in double cubic setting is given. The first step in space
group determination is to search for systematic absences in the list of
reflections. We observe no intensity for the reflections shown in
Table~\ref{tbl:extinctions}. Note that the indices are in double cubic setting
and therefore allow cyclic permutation, {\it i.e.} $h00$, $0k0$ and $00l$ are
all represented by $h00$. These reflection conditions only allow the space
groups $P2_13$ and $P4_232$. Attempts to determine the structure using these
space groups were unsuccessful. But, these space groups only have $h00$~:~$h=2n$
as reflection condition. This suggests that we have more information than can be
attributed to these space groups. Furthermore, the observed and, for cubic
systems, anomalous reflection conditions suggested that we might be looking at a
twinned crystal. In the next paragraph, we present the transformation of the
reflection conditions for orthorhombic $Pnma$ according to the presented twin
model.

\begin{table}
  \caption{Extinct reflections as observed in the
  $2a_{\textrm{p}}\times2a_{\textrm{p}}\times2a_{\textrm{p}}$ data set.}
  \label{tbl:extinctions}
  \begin{ruledtabular}
  \begin{tabular}{cllc}
 & Reflection & Extinction condition & Observed reflections\\
 & $h''$ $0$ $0$               & $h\neq2n$      & \textsc{e00}\cite{orientation}\\
 & $h''$ $\overline{h''}$ $0$  & $h\neq2n$      & \textsc{ee0}\cite{orientation}\\
 & $h''$ $k''$ $h''$           & $h\neq2n,k=2n$ & \textsc{eoe}\cite{orientation}\\
 & $h''$ $k''$ $\overline{h''}$& $h\neq2n,k=2n$ & \textsc{eoe}\cite{orientation}\\
  \end{tabular}
  \end{ruledtabular}
\end{table}

The following equations are used to transform the simple cubic Miller
indices $hkl$ to the standard $Pnma$ setting and to transform these to
the double cubic setting. 
    \begin{eqnarray}
      h'k'l'=h+l,~2k,~h-l \label{eq:cub naar pnma} \\
      h''k''l''=h'+l',~2k',~h'-l' \label{eq:pnma naar dubcub}
    \end{eqnarray}

The reflection conditions for $Pnma$ are shown in Table~\ref{tbl:twin
Pnma reflection conditions}. Using Eq.~\ref{eq:pnma naar dubcub}
the $Pnma$ reflection condition $0$~$k'$~$0$~:~$k'=2n$ is transformed
to double cubic as $0$~$k''$~$0$~:~$k''$~$=$~$2n$ and with cyclic
permutation yielding $h''$~$0$~$0$~:~$h''=2n$ and
$0$~$0$~$l''$~:~$l''=2n$. This corresponds to the reflection condition
$h''$~$0$~$0$~:~$h''=2n$ in Table~\ref{tbl:extinctions}.  However,
$Pnma$ reflections of the type $h'$~$0$~$h'$ are also transformed to
$h''$~$0$~$0$, with $h''=2h'$. Note that the orthorhombic to double
cubic transformation gives $0$~$k'$~$0=h''$~$0$~$0$, with $b'\parallel
a''$, and the reflection condition $k'=2n$ implies that only $h''=2n$
are present. While the $h'$~$0$~$h'$ reflections intrinsically give
$h''$~$0$~$0$ with $h''=2n$, with $b'\parallel b''$. This implies that
$h'$~$0$~$h'$ may never contribute intensity to reflections
$h''$~$0$~$0$, with $h''$ odd. Conversely, if we do not observe
intensity on reflections $h''$~$0$~$0$~:~$h''\neq2n$, then all of the
constituting $Pnma$ reflections, at most six, should be absent.

\begin{table}
   \caption{Reflection conditions for space group $Pnma$.}
   \label{tbl:twin Pnma reflection conditions}
   \begin{ruledtabular}
   \begin{tabular}{clllc}
     &  $0kl$ &:& $k$+$l=2n$ & \\
     &  $0k0$ &:& $k=2n$ & \\
     &  $00l$ &:& $l=2n$ & \\
     &  $hk0$ &:& $h=2n$ & \\
     &  $h00$ &:& $h=2n$ & \\
   \end{tabular}
   \end{ruledtabular}
\end{table}

To transform and understand the other reflection conditions of $Pnma$ we have to
take into account that an orthorhombic, and also a cubic, unit cell implies that
$hkl$ is equivalent upon sign reversal of each of the Miller indices. The $Pnma$
reflection condition $h'$~$0$~$0$~:~$h'$~$=$~$2n$ becomes
$h''$~$0$~$h''$~:~$h''=2n$ and similarly $0$~$0$~$l'$ transforms to
$l''$~$0$~$\overline{l''}$. But as we just stated, $h''$~$0$~$h''$ is identical
to $h''$~$0$~$\overline{h''}$. If we combine reflections then a reflection condition
is fulfilled if any of the contributing parts has intensity. To obtain
extinction, all contributing reflections should be extinct. As $0$~$0$~$l'$ has
no extinction condition, {\it i.e.} intensity for all $l'$, they will contribute
to all $h''$~$0$~$h''$~/~$ h''$~$0$~$\overline{h''}$ reflections, and the $Pnma$
reflection condition $h'$~$0$~$0$~:~$h'=2n$ is masked.

The overlap of different reflections also occurs for $0$~$k'$~$h'$ and
$h'$~$k'$~$0$. These transform to $h''$~$k''$~$\overline{h''}$, and
$h''$~$k''$~$h''$. Therefore, we cannot disentangle $0$~$k'$~$h'$ and $h'$~$k'$~$0$ in
the measured double cubic cell. The $Pnma$ reflection conditions for
$0$~$k'$~$l'$ and $h'$~$k'$~$0$ are $k'$+$l'=2n$ and $h'=2n$, respectively.
$h''$~$k''$~$h''$ and $h''$~$k''$~$\overline{h''}$ are found to be extinct for
$h''\neq2n$ and $k''=2n$ in the double cubic setting. To have these reflection
extinct, all contributing $Pnma$ reflection conditions must be extinct. This
implies that \emph{both} $h'=2n$ \emph{and} $k'$+$h'=2n$, which are the
reflection conditions for the contributing $0$~$k'$~$h'$ and $h'$~$k'$~$0$, may
not be fulfilled. Thus, taking into account that $h''=h'$+$0$ and $k''=k'$, it
is easily derived that $h''\neq2n$ and this yields $k''=2n$. This means
\textsc{oeo} reflections are extinct.

Now we consider reflections of the type $h''$~$k''$~$h''$ that are not extinct.
Our twin model yields six fractions that can contribute to the considered
reflections.

\begin{enumerate}
\item
$h''=2n$ and $k''=2n$. Reflections \textsc{eee} have contributions from the four
fractions with the doubled $b'$~axis parallel to either $a''$ or $c''$, this is
also true for reflections \textsc{ooo}. In addition, we have contributions from
the two fractions with $b'$ parallel to $b''$, \emph{i.e.} from $h'$~$k'$~$0$
and $0$~$k'$~$h'$.
\item
$h''\neq2n$ and $k''\neq2n$. Reflections \textsc{ooo}, in addition to the four
fractions with the doubled $b'$~axis parallel to either $a''$ or $c''$, have one
additional contribution from $0$~$k'$~$h'$. In the special case that $h''=k''$,
for both \textsc{eee} and \textsc{ooo}, then two of the four fractions, parallel
to $a''$ and $c''$ contribute.
\item
$h''=2n$ and $k''\neq2n$. Reflections \textsc{eoe}, only have a contribution
from the $h'$~$k'$~$0$ reflection, with $b'$ parallel to the observed
$b''$~axis.
\end{enumerate}

We have shown that the intensity of observed $h''$~$k''$~$h''$ reflections
depends on odd or even. We conclude that $h''$~$k''$~$h''$ reflections are
extinct for \textsc{oeo}, while the other reflections have contributions from up
to six twin fractions.

\label{app:intensity}

We can also learn something about the twinned origin by studying the intensity
distribution. Our twin model provides a natural explanation for the hierarchy in
intensities of the reflections. First, we consider the known transition from the
single cubic unit cell to the orthorhombic one to elucidate the patterns that
are inherent to perovskites. Then we proceed to show how these patterns are
convoluted in the double cubic, twinned model.

\begin{figure}
   \centering
   \includegraphics[bb=25 680 280 820]{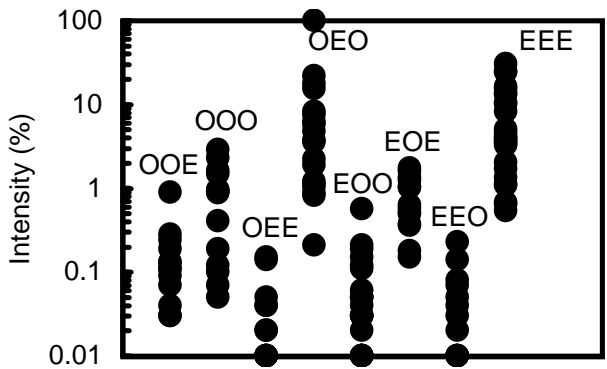}
   \caption{The calculated intensities for a conventional perovskite, subdivided
   with respect to odd and even of the Miller indices. $k=2n$ reflections are
   strongest, if $h'+l'=2n$. Reflections with $h'+l'\neq2n$ have roughly 100
   times less intensity.}
   \label{fig:Pnma}
\end{figure}

The intensity distribution of a normal $Pnma$ perovskite is shown in
Fig.~\ref{fig:Pnma}. In conventional orthorhombic $Pnma$ the reflections
$h'$~$k'$~$l'$ with $h'$+$l'=2n$ are stronger than $h'$+$l'\neq2n$. From these,
the $k=2n$ reflections are stronger than the $k\neq2n$ ones. This can be
attributed to their origin in the 3.9 \AA\ cubic structure. The $h'$~$k'$~$l'$
reflections in orthorhombic setting originate from $hkl$ in the single cubic
cell according to Eq.~(\ref{eq:cub naar pnma}). Therefore, $h'$ and $l'$ will
always be both even or both odd if the reflection {$h'$~$k'$~$l'$} is related to
an integer single cubic crystal plane. In other words if {$h'$-$l'\neq2n$} then
the reflections originate from a crystal plane in the single cubic setting with
a non-integer Miller index, \emph{i.e.} a superlattice reflection and their
intensities are typically weak. The same argument can be applied to $k$. As
$k'=2k$, reflections with $k'=2n$ originate from a regular crystal plane and
reflections with $k'\neq2n$ are superlattice reflections. These reflections with
$k'\neq2n$ however appear to have somewhat more intensity than the
$h'$-$l'\neq2n$ reflections.

We observe the following patterns in the intensity distribution of the double
cubic cell. First, reflections \textsc{eee} are generally the strongest,
\textsc{eoo}\cite{orientation} usually absent, or very weak. The following
hierarchy can be made if we disregard the cubic symmetry. \textsc{eoe}
reflections were much stronger than \textsc{eeo}, which are again stronger than
\textsc{oee}. Likewise, \textsc{oeo} reflections were, although weak, stronger
than \textsc{ooe} and again stronger than \textsc{eoo}.

In conventional $Pnma$ structure we found that if $h'$+$l'\neq2n$ then
reflections with $k'\neq2n$ are stronger than those with $k'=2n$. In our
even/odd notation, for the $Pnma$ structure this yields
(\textsc{eee}=\textsc{oeo}) $>>$ (\textsc{ooo}=\textsc{eoe}) $>>$
(\textsc{ooe}=\textsc{eoo}) $>$ (\textsc{oee}=\textsc{eeo}). So there are
roughly four intensity groups. We can correlate these four groups with the four
types of reflections we measured in the super cubic setting, with the following
intensity hierarchy:
\textsc{eee}$>>$\textsc{eeo}$>>$\textsc{ooo}$>$\textsc{eoo}.

Here we consider the transformation from the orthorhombic to the double cubic
setting for the in-plane indices, $h'$ and $l'$. Double cubic indices are
calculated with $h'$+$l'$ and $h'$-$l'$, therefore $h''$ and $l''$ will be both
even or both odd. Reflections of the type \textsc{e?o} or \textsc{o?e} will have
no intensity of the twin with $b'$ parallel to $b''$, as they are not
originating from an 'integer' crystal plane. In contrast with the single cubic
to orthorhombic $Pnma$ transition, where orthorhombic superlattice reflections
are allowed, the orthorhombic to double cubic transformation is just a new
choice of reference system, that may not and cannot generate new reflections.
But we have a mixing of different orientations of the $b'$~axis along the three
cubic axes $a''$, $b''$ and $c''$, which do generate reflections where none are
observed without the twinning. We can cycle the indices of these reflections so
that we get either \textsc{e?e} or \textsc{o?o}. From the corresponding twin,
with $b'$ parallel to either $a''$ or $c''$, we do have intensity on these
\textsc{e?o} and \textsc{o?e} reflections. \textsc{eoe} and \textsc{oeo} will
only have contributions from this particular setting with $b'\parallel b''$.
\textsc{eee} and \textsc{ooo} can be cycled and in general have contributions
from all six orientations.

\subsubsection{intensity}\label{app:distribution} The small deviations from the
cubic symmetry are reflected in the intensities of the reflections. To
illustrate this point, we transform an arbitrary, subcell, reflection $hkl$ to
the orthorhombic setting by Eq.~(\ref{eq:cub naar pnma}). Thus
orthorhombic reflections with $k'$ even and $h'$ and $l'$ both even or both odd
stem from planes that already existed in the original cubic unit cell. We
observe in all diffraction patterns of orthorhombic perovskites that reflections
that originate from cubic crystal planes generally have higher intensities than
those have that do not.

The intensity of a particular twin fraction is proportional to the volume of
that twin fraction. Using the uniqueness of some reflections, \emph{i.e.}
belonging to only one twin fraction lattice, we can search for the orientation
of the largest fraction. All $b'$~axes are oriented along one of the measured
double cubic axis. It is impossible to differentiate the two fractions with
parallel $b'$ and perpendicular $a'$ and $c'$. We considered groups of
reflections with cycled indices, having either one odd, \textsc{oee}, or one
even index, \textsc{eoo}. Roughly, the intensities of the reflections within
these groups occurred with ratio 70:20:10. This suggested that one twin had 70\%
of the volume, the others 20\% and 10\%.

These reflections are sorted to find the corresponding orientations. The sort
parameter is the observed double cubic axis that corresponded with the odd
indices in \textsc{oee}, and the even indices in \textsc{eoo}. Only cycled
reflections that occurred three times with measurable intensity were taking into
account. We found that $I_{a''}:I_{b''}:I_{c''}=5:80:15$. The constraint that
all three intensities have positive values, ignores the weakest reflections.
This method also ignores Bijvoet pairs, as we only measured $-20<l''<2$. In
principal \textsc{eoo} and $\overline{\textsc{eoo}}$ should have identical
intensity. If we indeed had a cubic system then these variations should be zero
within standard deviation. The distribution of the intensities, with respect to
the different axes, strongly suggests that the crystal is twinned.

\subsection{Intensity distribution}\label{subsec:Intensity distribution}
Careful analysis of the intensity distribution shows extra evidence for the
proposed models. Two important observations are made. First, the full data set
shows that reflections $hkl$ of type \textsc{eee} are by far the strongest.
Secondly, reflections having one odd Miller index, \textsc{oee}, \textsc{eoe}
nband \textsc{eeo}, are second strongest in intensity. Of these reflections
\textsc{eoe} has the highest intensity as seen in Fig.~\ref{fig:eeo}.

To understand the observed intensity distribution in the
$2a_p\times2a_p\times2a_p$ unit cell, we examine the influence of the structural
deformations on the structure factors. Despite the effect of Jahn-Teller
distortions, GdFeO$_3$ rotations and twinning the structure remains in origin
cubic. This means that the intensity distribution of the peaks with high
intensity will always mimic the intensity distribution of the peaks of the
undistorted cubic perovskite. An arbitrary reflection, in the small cubic
subcell, $hkl$, is transformed to the orthorhombic unit cell using equation{\ref{eq:cub naar pnma}}
Thus orthorhombic reflections with $h'$ and $l'$ both even or both odd and $k'$ even stem from the
original cubic unit cell. These reflections therefore have the highest
intensity, see Fig.~\ref{fig:Pnma}. Although the fact that the
$h'$~$k'$~$l'=$~\textsc{eoe} reflections do not originate from the original
cubic planes, they are allowed in the $Pnma$ symmetry. In Table~\ref{tbl:twin
Pnma reflection conditions} the reflection conditions for the $Pnma$ space group
are given. Extinctions observed in the double cubic setting are thoroughly
discussed before. Reflections $h'$~$k'$~$l'$ that
satisfy the condition $h'$~$k'$~$l'=$~\textsc{eoe} have higher intensities than
those that do not.

With these observations, we return to the observed double cubic lattice. To
double the $Pnma$ unit cell to a $2a_p\times2a_p\times2a_p$ unit cell, a
reflection $h'$~$k'$~$l'$ is transformed to $h''$~$k''$~$l''$ using Eq.~(\ref{eq:pnma naar dubcub}). The transformation yields that double cubic
reflections with $k''=2n$, $k''=k'$, originate from single cubic reflections
with $k=n$. Likewise, $k''=2n+1$ reflections originate from non-integer single
cubic indices $k$, and will therefore have less intensity. This explains why the
measured $h''k''l''=$\textsc{eee} reflections are strongest and \textsc{eoe} are
one order of magnitude less strong.

We have seen that $h''=h'$+$l'$ and $l''=h'$-$l'$ thus $h''$ and $l''$ are
always both even or both odd. Therefore \textsc{eoe} only has contributions of
the fractions with $k''$ corresponding to the orthorhombic $b$~axis.
\textsc{oee} and \textsc{eeo} originate from fractions with their $b$~axis
parallel to $a''$ and $c''$, respectively. The fact that our observed
\textsc{eoe} is stronger than \textsc{oee} and \textsc{eeo}, indicates that, by
chance, the observed $b''$~axis is parallel with the $b'$~axis
of the largest twin fraction.

\subsection{Refinement} Both the Ca content and the volume fractions of the
twins can be determined from single crystal x-ray diffraction in the following
way. First a full hemisphere of reciprocal $hkl$ space is measured. On this very
large data set, the atomic positions, the Ca content and the volume fractions
are refined. The results of the refinement are compared with a refinement on
only one octant of these measured reflections. The outcome was equivalent within
the error bars. We refined again fixing the twin fractions and Ca concentration
to the values found for the first refinement on the large data set, and allowing
only the atomic co-ordinates and anisotropic displacement parameters to change.
This resulted in the same structural model as found with the complete structure
determination on the largest data set.

It is concluded that apparently the refinement is insensitive for the size of
the data set. However, it will be dependent on the $\theta$~range of the data
set. The insensitivity can be understood if we view the symmetry relations and
Friedel pairs. Due to the Patterson symmetry, $mmm$, the relation
$F(\overline{h}kl)=F(hkl)$ is valid for all structure factors. Friedel pairs are
reflections that have intrinsically the same structure factor, related thus by
symmetry and ignoring absorption and anomalous scattering effects. Although we
have six different orientations, one octant chosen in a particular orientation
will still contain data of all fractions. The six orientations can be
transformed in such a way that every $h'$~$k'$~$l'$ is represented for all three
$b$~directions in one octant of reciprocal space.

\section{Implementation in refinement}\label{app:model}
The standard refining programs can work with twin models, but only if they
consist of merohedral twins. In our case, the reflections do not correspond with
an orthorhombic unit cell, but with a, twice as large, pseudo cubic unit cell.
Here we will explain the way we worked around the refining program.

First, a hemisphere in $hkl$ space was measured with -$20<h<20$, -$20<k<20$ and
$0<l<20$. On this data set, we could refine our twin model with the six
fractions, including the Ca concentration on the A~site. Measuring such a large
range in $hkl$ space requires a large amount of time, roughly 10 days in the
present set-up. Typically, for crystals with an orthorhombic unit cell measuring
one octant of reciprocal space is sufficient. From refinements on selected parts
of the data set, we concluded that we could investigate the structure, for
instance at different temperatures, by measuring only the positive octant of the
main fraction, thereby considerably limiting the measuring time.

Here we describe the model and the application by SHELXL. The standard refining
programs can work with merohedral twin models. However, in our case, the
observed lattice does not coincide with one orthorhombic unit cell. Every
observed reflection has contributions of up to six twin fractions. We measured
both in the double cubic setting as in the orthorhombic setting of the main twin
fraction. We transformed the $h''k''l''$ indices to the six possible twin
orientations. Three of the possibilities are given by
\begin{eqnarray}
   h'k'l'=\frac{1}{2}(h''+l''), k'', \frac{1}{2}(h''-l'') \\
   h'k'l'=\frac{1}{2}(l''+k''), h'', \frac{1}{2}(l''-k'') \\
   h'k'l'=\frac{1}{2}(k''+h''), l'', \frac{1}{2}(k''-h'')
\end{eqnarray}
The other three can be acquired by changing $h'$ for $l'$ and $l'$ for -$h'$.

A~new software program TWINSXL was developed to transform the standard data
file, HKLS, by using the appropriate transformation matrices from a second input
file\cite{Mee00}. The new data file is constituted of lines of $h, k, l$,
intensity plus standard deviation and twin fraction number. We used the "HKLF 5"
option of SHELXL to refine data. The refinement uses a crystal model for the
orthorhombic structure with the normal, adjustable variables. Five variables for
the twin fractions were added, the sixth fraction is calculated as the
complement of the other five fractions. The sum of the appropriate calculated
intensities for all fractions was compared with the observed integrated
intensities.



\begin{thebibliography}{33}
\bibitem[{gla()}]{glazer}
\bibinfo{note}{$p^-q^+r^0$ means that along [100] a rotation of size $p$ is
  alternated positive and negative, along [010] a rotation of size $q$ is
  always (arbitrarily) positive, and along [001] there is no rotation.}
\bibitem[{pos()}]{position}
\bibinfo{note}{The O2 position, ($x_{O2}, y_{O2}, z_{O2}$) near ($\frac{1}{4},
  0, \frac{1}{4}$) is used here. The other three oxygen positions in the same
  \emph{xz} plane of the unit cell follow from the space group symmetry. For
  these the shifts are also transformed according to the symmetry operations.}
\bibitem[{ori()}]{orientation}
\bibinfo{note}{We do not explicitly mention the different orientations of the
  orthorhombic unit cell. If the regarded $b$~axis is not parallel with $b''$,
  than the indices should be cycled appropriately.}
\end{thebibliography}
\end{document}